\pdfoutput=1
\documentclass[superscriptaddress,aps,twocolumn,floatfix,prb]{revtex4}
\usepackage{graphicx,amsmath,bbm}

\newcommand{\vk}{\vec k}
\newcommand{\ve}{\vec e}

\newcommand{\vq}{\vec q}

\renewcommand{\vr}{\vec r}

\newcommand{\vs}{\vec s}

\newcommand{\vsigma}{\mbox{\boldmath $\sigma$}}

\renewcommand{\Im}{\mbox{Im}\,}

\renewcommand{\vec}[1]{\mathbf{#1}}

\newcommand{\ua}{\uparrow}
\newcommand{\da}{\downarrow}

\newcommand{\SWgap}{\Delta^{\rm sw}}

\begin{document}

\title{Spin-wave-induced corrections to the electronic density of states\\ in metallic ferromagnets}
\date{\today}

\author{A.\ Ricottone}
\affiliation{Dahlem Center for Complex Quantum Systems and Fachbereich Physik, Freie Universit\"{a}t Berlin, Arnimallee 14, 14195 Berlin, Germany}
\author{J.\ Danon}
\affiliation{Dahlem Center for Complex Quantum Systems and Fachbereich Physik, Freie Universit\"{a}t Berlin, Arnimallee 14, 14195 Berlin, Germany}
\affiliation{Niels Bohr International Academy, Niels Bohr Institute, University of Copenhagen, Blegdamsvej 17, 2100 Copenhagen, Denmark}
\author{P.~W.\ Brouwer}
\affiliation{Dahlem Center for Complex Quantum Systems and Fachbereich Physik, Freie Universit\"{a}t Berlin, Arnimallee 14, 14195 Berlin, Germany}

\begin{abstract}
We calculate the correction to the electronic density of states in a disordered ferromagnetic metal induced by spin-wave mediated interaction between the electrons. Our calculation is valid for the case that the exchange splitting $\Delta$ in the ferromagnet is much smaller than the Fermi energy, but we make no assumption on the relative magnitude of $\Delta$ and the elastic electronic scattering time $\tau_{\rm el}$. In the ``clean limit'' $\Delta \tau_{\rm el}/\hbar \gg 1$ we find a correction with a $T^{d/2}$ temperature dependence, where $d$ is the effective dimensionality of the ferromagnet. In the ``dirty limit'' $\Delta \tau_{\rm el}/\hbar \ll 1$, the density-of-states correction is a non-monotonous function of energy and temperature.
\end{abstract}

\maketitle

\section{Introduction}\label{sec:intr}

Whereas quantum corrections to the electronic properties of normal metals have been studied theoretically and experimentally for almost half a century,\cite{Beenakker19911,aa:book,imry:book,akkermans} the electronic properties of ferromagnetic metals only attracted attention at a later stage, mainly triggered by the emergence of the field of spintronics.\cite{spinrev} Conceptually simple devices, such as a ferromagnet--insulator--ferromagnet junction, were found to exhibit spectacular magnetoresistive effects.\cite{Miyazaki1995L231,PhysRevLett.74.3273} Ferromagnet--superconductor junctions display a wide range of interesting phenomena, including inhomogeneous induced superconductivity,\cite{PhysRevLett.86.304} polarization-dependent Andreev scattering,\cite{Soulen02101998} and induced triplet superconductivity,\cite{keizer:nat} all stemming from the different spin-orderings in ferromagnets and ($s$-wave) superconductors. In the context of these fairly recent discoveries, a detailed understanding of the electronic properties of disordered ferromagnets is quintessential.

Although the two key elements determining quantum corrections in normal metals, disorder and Coulomb interactions, also play a role in ferromagnets, the existence magnetic of order in ferromagnets adds additional complexity. The very presence of the magnetic order, but also its quantum and thermal fluctuations, modify quantum corrections to electronic properties in an essential way. For example, spin-orbit interaction couples the orbital motion of electrons to the exchange field of the magnet, causing an anomalously strong dependence of the conductivity on magnetization direction.\cite{PhysRevB.64.144423,kn:adam2006b} Also electronic scattering from spatial or temporal inhomogeneities of the magnetization in space and time, such as domain walls and spin waves, can lead to dephasing\cite{PhysRevLett.81.3215,PhysRevB.84.224433} and it can qualitatively affect the metallic resistivity.\cite{PhysRevLett.78.3773,PhysRevLett.79.5110,PhysRevB.79.140408} 

In this article we address the tunneling density of states, which is a key property in devices in which metallic and insulating layers alternate. For a normal metal, the interplay of disorder and Coulomb interaction effects causes the tunneling density of states to deviate from the thermodynamic density of states for low temperatures and excitation energies.\cite{Altshuler1979115,PhysRevLett.44.1288,Altshuler1979968,aa:book} In three dimensions, the temperature dependence of the correction to the density of states follows a square-root power law. While it is clear that for a weakly disordered ferromagnet the same effects modify electronic density of states, in a ferromagnet fluctuations of the magnetic order may cause an additional correction to the density of states.


About a decade ago, a series of experiments on ferromagnetic tunnel junctions\cite{Moodera1999248} revealed a $T^{3/2}$ power law up to room temperature of the tunnel resistance and magnetoresistance of the junctions.\cite{PhysRevB.58.R2917,5028809} MacDonald {\em et al.}\cite{PhysRevLett.81.705} linked this observation to a reduction in the average (surface) polarization of the magnets, caused by thermally excited spin waves, which is indeed known to have a $T^{3/2}$ dependence.\cite{Mills19671855} Essentially, the mechanism of Ref.\ \onlinecite{PhysRevLett.81.705} is that thermal fluctuations of the magnetization, which in Ref.\ \onlinecite{PhysRevLett.81.705} is assumed to be fully carried by the conduction electrons, smear the difference between majority and minority electron densities of states.

In this article, we revisit the question of a spin-wave-mediated correction to the density of states. 
We take the viewpoint that the emission and absorption of spin waves leads to an effective interaction between the electrons (in the same way as that the emission and absorption of phonons leads to an effective electronic interaction in the theory of superconductivity), and consider the resulting correction to the density of states for a disordered ferromagnet, analogous to the Coulomb-interaction induced correction to the density of states of a normal metal. Remarkably, in the most relevant parameter range we find the same $T^{3/2}$ dependence of the correction to the density of states as in Ref.\ \onlinecite{PhysRevLett.81.705}, although our calculation does not rely on different (bare) densities of states for majority and minority electrons. We note that quantum corrections to the conductivity and the dephasing rate that arise from an effective spin-wave-mediated electron-electron interaction have been considered previously by various authors.\cite{JPSJ.72.1155,PhysRevB.79.140408,muttalibwoelfle} However, we are not aware of a calculation of the effect on the tunneling density of states.

Our calculations are performed using diagrammatic perturbation theory. A key condition for applicability of this approach is that the exchange splitting between majority and minority spins $\Delta$ be much smaller than the Fermi energy $E_{\rm F}$. We note that this condition is not met for strong ferromagnets, such as Fe or Ni, for which $\Delta$ and $E_{\rm F}$ differ less than one order of magnitude. Nevertheless, we point out that our calculation provides a controlled theoretical estimation of the effect of the effective spin-wave-mediated interaction, as it identifies temperature and energy dependencies of the density of states that are unrelated to a difference in the bare densities of majority and minority electrons. Our calculations do not rely on the diffusion approximation, so that $\Delta$ may be large or small in comparison to the elastic scattering rate $\hbar/\tau_{\rm el}$. This is important, because most realistic ferromagnets are in the ``clean limit'', in which $\Delta \tau_{\rm el}/\hbar \gtrsim 1$. 

The precise microscopic model we consider is described in Sec.\ \ref{sec:model}. The calculation of the leading correction to the density of states using diagrammatic perturbation theory is given in Sec.\ \ref{sec:dos}, followed in Sec.\ \ref{sec:4} by a discussion of the result in the limits of a clean and a dirty ferromagnet ($\Delta$ large or small in comparison to $\hbar/\tau_{\rm el}$, respectively). Our main result is a clean-limit density-of-states correction proportional to $T^{d/2}$ or $|\epsilon|^{d/2}$ in an effectively $d$-dimensional ferromagnet, where $\epsilon$ is the excitation energy (measured with respect to the Fermi energy). In the dirty limit we find a non-monotonous energy and temperature dependence. We conclude with a comparison to the density of states from Coulomb interactions in Sec.\ \ref{sec:6}.

\section{Model}\label{sec:model}

To describe the conduction electrons in the disordered ferromagnet and their interaction with fluctuations of the magnetization of the $d$-band electrons, we use the same model as employed in Ref.\ \onlinecite{PhysRevB.84.224433}. An identical model description has been used for ferromagnetic metals in which the magnetism resides with itinerant electrons only.\cite{PhysRevLett.81.705} In this model, the electrons are described with the effective single-particle Hamiltonian
\begin{equation}
H = \frac{\hbar^2k^2}{2m} - \mu + V(\vr) - J\, \vs(\vr) \cdot \vsigma,
\label{eq:ham1}
\end{equation}
where the first term represents the kinetic energy, $\mu$ is the chemical potential, $V(\vr)$ is the impurity potential, and the last term describes the exchange interaction between the conduction electrons and the $d$-band electron spins. In the exchange term, $J$ is the exchange constant and $\hbar \vs(\vr)$ is the spin density of the $d$-electrons. 

We choose the $z$-axis such that it points in the direction of the mean $d$-band magnetization, $\bar{\vs} = \bar{s} \ve_z$ and we write
\begin{equation}
 -J \,\vs(\vr) \cdot \vsigma = 
- \frac{1}{2}\Delta\sigma_z
+  H_{sd,{\rm fluc}}.
\end{equation}
The mean magnetization gives rise to an effective exchange splitting $\Delta = 2 J \overline{s}$ and the coupling to the fluctuations around this mean value are described by $H_{sd,{\rm fluc}}$.\cite{ferro_dos:footnote1} To linear order in the fluctuations we can focus on the transverse components $s_{x,y}(\vr,t)$,
\begin{equation}
H_{sd,{\rm fluc}} =  -J \left( \begin{array}{cc} 0 & s_-({\bf r},t) \\ s_+({\bf r},t) & 0 \end{array} \right),
\label{eq:hamsd}
\end{equation}
where we use the notation $s_{\pm} = s_x \pm i s_y$. 

Dynamical processes involving the absorption and excitation of a $d$-band spin wave are characterized by the transverse spin susceptibility
\begin{equation}
  \chi^{\rm R}_{-+}(\vr-\vr',\tau) =
  - i \Theta(\tau) \langle [ s_-(\vr,\tau), s_+(\vr',0) ] \rangle,
  \label{eq:spincorr}
\end{equation}
where $\Theta(\tau) = 1$ for $\tau > 0$ and $\Theta(\tau) = 0$ otherwise is the Heaviside step function. The susceptibility $\chi^{\rm R}_{-+}(\vr,\tau)$ describes the response of the $d$-electron spin density to an applied magnetic field. Its Fourier transform $\chi^{\rm R}_{-+}(\vq,\omega)$ is conveniently expressed in terms of the spin wave frequencies $\omega_{\vq}^{\rm sw} = \omega_{-\vq}^{\rm sw}$,\cite{kubospinsus,JPSJ.72.1155}
\begin{eqnarray}
  \chi^{\rm R}_{-+}(\vq,\omega) &=&
  \int d\vr \int d\tau \, \chi^{\rm R}_{-+}(\vr,\tau) 
  e^{i \omega \tau-i \vq \cdot \vr } \\ &=&
  \frac{ - 2 \bar{s}}{\omega + \omega_{\vq}^{\rm sw} + i\eta},
  \label{eq:chisw}
\end{eqnarray}
where $\eta$ is a positive infinitesimal. The susceptibility for opposite spin orientations reads
\begin{equation}
  \chi^{\rm R}_{+-}(\vq,\omega) =
  \frac{2 \bar{s}}{\omega - \omega_{\vq}^{\rm sw} + i\eta}.
\end{equation}

The spin wave frequencies $\omega_{\vq}^{\rm sw}$ are determined by interactions and anisotropy factors not taken into account in the conduction electron Hamiltonian (\ref{eq:ham1}). Following Refs.\ \onlinecite{PhysRevB.79.140408} and \onlinecite{muttalibwoelfle} we assume the phenomenological isotropic spin-wave dispersion relation
\begin{equation}
\hbar\omega_\vq^{\rm sw} = \hbar D^{\rm sw} q^2 + \SWgap.
\label{eq:disp}
\end{equation}
Here $D^\text{sw}$ is the spin wave stiffness, usually of the order $\hbar D^\text{sw} \sim \Delta/k_F^2$. The constant $\SWgap$ gives the spin wave gap, which can be due to, e.g., an externally applied magnetic field in the $z$-direction or by the magnetocrystalline anisotropy of the material. In the former case, one has $\SWgap = g\mu_B B$, where $B = B_{\rm ext} - g\mu_B \mu_0 \bar{s} \xi/4 \pi$ is the external magnetic field corrected for the demagnetizing field of the device, $\xi$ being a numerical constant determined by the shape of the ferromagnet.\cite{kittel} In the latter case $\SWgap = 2K/\bar{s}$, where $K$ is the energy density characterizing the anisotropy.

\section{Perturbative calculation}\label{sec:dos}

The emission and absorption of spin waves gives rise to an effective interaction between the electrons. For nonzero spin-wave gap this interaction is short-range, and its effect on the electronic density of states of a disordered ferromagnet can be calculated via diagrammatic perturbation theory. In order to apply the diagrammatic perturbation theory it is necessary that all relevant energy scales be small in comparison to the Fermi energy $E_{\rm F}$. In the present case, this means that the exchange splitting $\Delta$ and the elastic scattering rate $\hbar/\tau_{\rm el}$ must be small in comparison to $E_{\rm F}$. No assumptions need to be made with regard to the relative magnitude of $\Delta$ and $\hbar/\tau_{\rm el}$. The condition $\Delta \ll E_{\rm F}$ is not met (as a strong inequality) for the elemental ferromagnets, which means that for those materials our results should be seen as order-of-magnitude estimates.

The same condition $\Delta \ll E_{\rm F}$ implies that, without the spin-wave-mediated correction, majority and minority electrons have the same density of states $\nu$ at the Fermi level, the same Fermi velocity $v_{\rm F}$, and the same elastic scattering time $\tau_{\rm el}$. 
(As shown below, we will find that for the clean limit $\Delta\tau_{\rm el}/\hbar \gg 1$ the corrections do not depend on $\tau_{\rm el}$, rendering it thus unnecessary to keep track of two different scattering times.) 
Consistent with these expectations, for the impurity potential we take a Gaussian white noise distribution,
\begin{equation}
  \langle V(\vr) V(\vr') \rangle = \frac{\hbar}{2 \pi \nu \tau_{\rm el}} \delta(\vr-\vr'),
\end{equation}
where $\nu$ is the conduction electron density of states (per spin direction) and $\tau_{\rm el}$ the elastic mean free time. 

\begin{figure}[t]
\begin{center}
\includegraphics{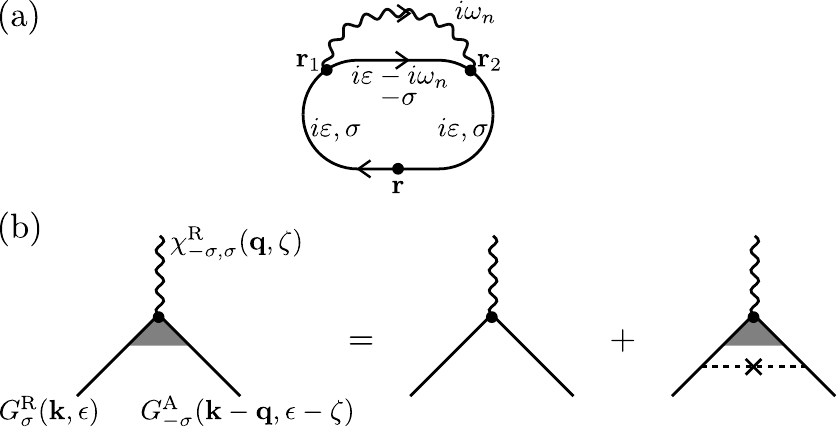}
\end{center}
\caption{(a) Diagram representing the leading-order correction to the density of states. Solid lines denote the conduction electron Green functions, the wiggly line represents the spin-wave propagator $\chi_{-\sigma,\sigma}$. (b) Dressed interaction vertex for a disordered ferromagnet. The dashed line with the cross represents correlated elastic impurity scattering.\label{fig:0}}
\end{figure}

In diagrammatic perturbation theory, the leading correction to the density of states $\nu_{\sigma}$ for electrons with spin $\sigma = \pm 1$ is given by the ``Fock'' diagram shown in Fig.\ \ref{fig:0}a. The ``Hartree'' correction is absent because of the spin-flip nature of the $sd$ interaction term. Expressing this correction in terms of the exact retarded and advanced Green functions $G^{\rm R}_{\sigma}(\vr,\vr',\epsilon)$ and $G^{\rm A}_{\sigma}(\vr,\vr',\epsilon)$ of the conduction electrons, and using the identity 
$$
  \int d\vr \, G^{\rm R}_{\sigma}(\vr_1,\vr,\epsilon) G^{\rm R}_{\sigma}(\vr,\vr_2,\epsilon) = - \frac{\partial}{\partial \epsilon} 
  G^{\rm R}_{\sigma}(\vr_1,\vr_2,\epsilon),
$$%
one finds
\begin{widetext}
\begin{eqnarray}
  \delta \nu_{\sigma} (\epsilon,T) &=&
  - \frac{J^2}{\hbar\pi {\cal V}} \mbox{Im}\,
  \int d\vr_1 d\vr_2 
  \left[ \frac{\partial}{\partial\epsilon}
  G_{\sigma}^{\rm R}(\vr_1,\vr_2,\epsilon)
  \right]
  \int \frac{d\zeta}{4 \pi i} 
  \left\{ 2 i \coth \Big(\frac{\zeta}{2 T} \Big)
  \vphantom{\frac{-1}{2}}
  G^{\rm R}_{-\sigma}(\vr_2,\vr_1,\epsilon-\zeta)
  \mbox{Im}\, \chi^{\rm R}_{-\sigma,\sigma}(\vr_2-\vr_1,\zeta/\hbar)
  \right. \nonumber \\ && \left. \mbox{}
  + \tanh \Big(\frac{\epsilon-\zeta}{2T} \Big)
  [G^{\rm R}_{-\sigma}(\vr_2,\vr_1,\epsilon-\zeta) - 
  G^{\rm A}_{-\sigma}(\vr_2,\vr_1,\epsilon-\zeta)] 
  \chi_{-\sigma,\sigma}^{\rm R}(\vr_2-\vr_1,\zeta/\hbar) \right\},
  \label{eq:deltanu}
\end{eqnarray}
\end{widetext}
where ${\cal V}$ is the sample volume.
In order to ensure convergence of the integration at large energies $\zeta$, we later subtract the zero-energy zero-temperature correction $\delta \nu_{\sigma}(0,0)$ from the expression for $\delta \nu_{\sigma}(\epsilon,T)$ shown above.

It remains to average the products of the electronic Green functions over the disorder. For the two contributions to $\delta \nu(\epsilon,T)$ that contain a product of two retarded Green functions, we may replace the product of the Green functions by the product of the ensemble-averaged Green functions $\langle G^{\rm R}_{-\sigma}(\vr_2,\vr_1,\epsilon) \rangle = \langle G^{\rm R}_{-\sigma}(\vr_2-\vr_1,\epsilon) \rangle$. Changing to the Fourier representation, the density of states correction can be expressed as a summation over spin-wave wave vectors $\vq$. One then quickly finds that these two contributions vanish, as long as the relevant wave numbers $q \ll k_{\rm F}$.

Such a procedure can not be applied to the term that has a product of a retarded Green function and an advanced Green function. Here impurity scattering renormalizes the two vertices for the electron-spin-wave interaction, see Fig.\ \ref{fig:0}b. 
The result is most conveniently expressed as
\begin{eqnarray}
\delta \nu_{\sigma} (\epsilon,T) &=& 
  \frac{2 J^2 \nu \tau_{\rm el}}{\hbar^2 {\cal V}}
  \Im\sum_{\vq}  \int \frac{d\epsilon'}{4\pi i} 
  \tanh \Big(\frac{\epsilon'}{2T} \Big)
  \\ && \mbox{} \nonumber  \times
  \chi^{\rm R}_{-\sigma,\sigma}[\vq,(\epsilon-\epsilon')/\hbar]
  \frac{\partial}{\partial\epsilon}
  \frac{1}{1-\Pi_{\sigma,-\sigma}(\vq,\epsilon,\epsilon')},
\end{eqnarray}
where the bare structure factor $\Pi_{\sigma,-\sigma}(\vq,\epsilon,\epsilon')$ is expressed in the disorder-averaged Green functions as
\begin{eqnarray}
  \Pi_{\sigma,-\sigma}(\vq,\epsilon,\epsilon') &=&
  \frac{\hbar}{2\pi\nu\tau_{\rm el} {\cal V}} 
  \label{eq:strfac}\\ && \mbox{} \times
  \sum_\vk
\langle G^{\rm R}_\sigma (\vk,\epsilon) \rangle \langle G^{\rm A}_{-\sigma} (\vk-\vq,\epsilon') \rangle.\nonumber 
\end{eqnarray}
Using the explicit expressions for the disorder-averaged Green functions,
\begin{align}
  \langle G^{\rm R}_\sigma (\vk,\epsilon) \rangle & = \langle G^{\rm A}_\sigma (\vk,\epsilon) \rangle^* =
  \frac{1}{\epsilon - \epsilon_{k} + \frac{\sigma}{2} \Delta + \frac{i \hbar}{2 \tau_{\rm el}}}, \label{eq:grav}
\end{align}
with $\epsilon_k = \hbar^2 k^2/2m - \mu$,
one finds that, as long as $\epsilon$, $\epsilon' \ll E_{\rm F}$, the structure factor depends on the energy difference $\epsilon-\epsilon'$ only, 

\begin{align}
\Pi_{\sigma,-\sigma}&(\vq,\epsilon-\epsilon') = \nonumber\\
& \frac{1}{2i q v_{\rm F}\tau_{\rm el}}
  \ln \frac{\hbar - i (\epsilon-\epsilon' + \sigma\Delta - \hbar q v_{\rm F})\tau_{\rm el}}
           {\hbar - i (\epsilon-\epsilon' + \sigma\Delta + \hbar q v_{\rm F})\tau_{\rm el}}.
  \label{eq:piresult}
\end{align}
Substituting Eq.\ (\ref{eq:chisw}) for the transverse spin susceptibility, one arrives at the final expression
\begin{eqnarray}
  \label{eq:dn}
  \frac{\delta \nu_{\sigma}(\epsilon,T)}{\nu} &=&
  - 
  \frac{4 \bar s J^2 \tau_{\rm el}}{\hbar {\cal V}} \mbox{Im}\,
  \sum_{\vq} \int \frac{d\zeta}{4 \pi i}
  \tanh \Big( \frac{\epsilon - \zeta}{2 T} \Big)
  \\ && \mbox{} \nonumber \times
  \frac{\sigma}{\zeta + \hbar \sigma \omega_{\vq}^{\rm sw} + i \eta}
  \frac{\partial}{\partial \zeta}
  \frac{1}{1 - \Pi_{\sigma,-\sigma}(\vq,\zeta)}.~~~
\end{eqnarray}

This expression, with Eq.\ (\ref{eq:piresult}) for $\Pi_{\sigma,-\sigma}$, presents the most general result obtainable within the perturbative approach and is the starting point for our further analysis. It does not rely on the diffusion approximation, i.e., on the smallness of $\Delta$, $\hbar \omega_{\vq}^{\rm sw}$, or $v_{\rm F} q$ with respect to $\hbar/\tau_{\rm el}$, and instead only assumes smallness with respect to $E_{\rm F}$. 

We note here that it follows from Eq.\ (\ref{eq:dn}) that 
\begin{equation}
  \delta \nu_\ua(\epsilon,T) = \delta \nu_\da(-\epsilon,T).
  \label{eq:symmetry}
\end{equation}
However, there is no symmetry relation relating the density-of-states corrections $\delta \nu_\ua$ and $\delta \nu_\da$ at equal energies, and in general $\delta \nu_\ua(\epsilon,T)$ will be different from $\delta \nu_\da(\epsilon,T)$ if $\epsilon \neq 0$. In our model, no difference between $\nu_\ua$ and $\nu_\da$ appears at zero energy.
Because of the symmetry (\ref{eq:symmetry}), we can focus on one particular spin direction, say spin up, without loss of generality, and the results for the other spin direction as well as the total correction $\delta \nu_{\ua} + \delta \nu_{\da}$ to the density of states simply follow from the symmetry relation (\ref{eq:symmetry}).

\section{Asymptotic behavior of the correction}
\label{sec:4}

Inserting the spin wave dispersion relation (\ref{eq:disp})---or, if desired, a more detailed dispersion including shape-dependent terms and various anisotropies\cite{kittel}---Eq.\ (\ref{eq:dn}) can be numerically integrated to give the energy and temperature dependence of the density-of-states correction for general values of the parameters. The contributions from the poles can be regularized by adding a small positive imaginary part to the integration variable $\zeta$, which preserves the causal dependencies of Eq.\ (\ref{eq:dn}). (Results will not depend on the value of the infinitesimal added.)

For the limiting case of a clean ferromagnet, $\Delta\tau_{\rm el}/\hbar \gg 1$, it is possible to simplify Eq.\ (\ref{eq:dn}) considerably and to arrive at analytic results. Below we will present the resulting corrections to the density of states in this clean limit for three-, two-, and one-dimensional samples, and we will also present numerical results for the opposite dirty limit where $\Delta\tau_{\rm el}/\hbar \ll 1$.

\subsection{Clean limit}

We first describe the case of a clean ferromagnet, for which the inequality $\Delta\tau_{\rm el}/\hbar \gg 1$ holds. 
If, in addition, we also assume that $\Delta \gg \hbar qv_{\rm F}$ for all $\vq$ which contribute significantly to the sum in (\ref{eq:dn}), the expression for the spin-wave correction to the density of states can be significantly simplified. Below we will show that this restriction is equivalent to assuming that $\Delta \gg E_{\rm F}^{2/3}{\rm max}\{T^{1/3},|\epsilon|^{1/3}\}$, which is not in contradiction with $\Delta \ll E_{\rm F}$ but limits the validity of our approach to temperatures and energies $T,|\epsilon| \ll \Delta^3/E_{\rm F}^2$.

Using the inequalities $\Delta \tau_{\rm el}/\hbar \gg 1$ and $\Delta \gg \hbar q v_{\rm F}$, one can expand (\ref{eq:dn}) for large $\Delta\tau_{\rm el}/\hbar$, yielding to leading order in $\hbar/\Delta\tau_{\rm el}$
\begin{align}
\frac{\delta \nu_{\ua} (\epsilon,T)}{\nu} = -\frac{J^2\bar{s}}{\Delta^2 {\cal V}}\sum_{\vq} \tanh \left(\frac{\epsilon+\hbar\omega_{\vq}^{\rm sw}}{2T} \right). \label{eq:nuclean}
\end{align}
This result does not depend on $\tau_{\rm el}$, which implies that all spin wave mediated electron-electron interactions are short-range and take place on a length scale much smaller than the elastic mean free path $l_{\rm el} = v_{\rm F} \tau$. Furthermore, we note that the apparent divergence of Eq.\ (\ref{eq:nuclean}) from the contribution from large wave vectors $q\to \infty$ is canceled if one calculates the difference with the zero-temperature zero-energy correction to the density of states, 
\begin{equation}
  \delta \nu^*_{\sigma}(\epsilon,T)\equiv \delta \nu_{\sigma}(\epsilon,T) - \delta \nu_{\sigma}(0,0),
\end{equation}
which gives
\begin{align}
  \frac{\delta \nu^*_{\ua} (\epsilon,T)}{\nu} = \frac{2J^2\bar{s}}{\Delta^2 {\cal V}}\sum_{\vq} n_{\rm F}(\epsilon+\hbar\omega_\vq^{\rm sw}), \label{eq:nusclean}
\end{align}
where $n_{\rm F}(\epsilon)$ is the Fermi function. 
Indeed, after subtracting $\delta \nu_{\ua}(0,0)$, all contributions coming from spin waves with an energy $\hbar\omega_{\vq}^\text{sw} \gtrsim {\rm max} \{ T,|\epsilon| \}$ are exponentially suppressed. Therefore, from the dispersion relation (\ref{eq:disp}) we see that the largest momenta which we have to take into account are of the order $$q_{\rm max} = {\rm max} \{ \sqrt{T/\hbar D^\text{sw}},\sqrt{|\epsilon|/\hbar D^\text{sw}}\}.$$ In order to satisfy $\Delta \gg \hbar qv_{\rm F}$ for all $\vq$ in the summation, we thus find the constraint $\Delta \gg  \hbar q_{\rm max} v_{\rm F}$. Using that typically $\hbar D^{\rm sw} \sim \Delta/k_{\rm F}^2$ we recover the constraint anticipated above. 

For ferromagnetic samples with large enough dimensions the summation over $\vq$ can be replaced by an integral which can be explicitly evaluated. If all three dimensions are much larger than $q^{-1}_{\rm max}$, we can treat the sample as three-dimensional and convert the sum over $\vq$ to an integral over spherical coordinates. If one or two of the dimensions are small enough, $a\ll q_{\rm max}^{-1}$ (but still $a \gg l_{\rm el}$), the effective dimensionality for the spin waves becomes lower, and the integral over $\vq$ becomes two- or one-dimensional as well. The constraint $a\ll q_{\rm max}^{-1}$ corresponds to the regime of low temperatures, where $T\ll \hbar D^{\rm sw}/a^2$. For example, for a thin Fe sheet or wire with thickness/diameter $a = 10$~nm, we find that a lower dimensional treatment is justified only if $T\ll 800$~mK using typical parameters for iron.\cite{ashmer,PhysRevLett.67.2363} The restrictions on the temperature are however not always that severe. For instance, for the thin Gd films studied in Ref.\ \onlinecite{PhysRevB.79.140408} and using their estimates for the material parameters, we find that the samples are effectively two-dimensional up to temperatures of a few times $10$~K. We note here that we take the system size in all dimensions to be larger than the electronic elastic scattering length, so that electronic transport is diffusive in all directions.

In keeping with the effective $d$-dimensional description, we introduce $d$-dimensional exchange constants and magnetization densities, by replacing $J \to J/a^{3-d}$ and $\bar{s} \to \bar{s}a^{3-d}$. Hence, $J$ and $\bar{s}$ now have respectively dimensions of energy times volume, area, or length and polarization per volume, area, or length. We then find
\begin{equation}
\frac{\delta \nu^*_{\ua} (\epsilon,T)}{\nu} = -\frac{J^2\bar{s}}{2^{d-1}\Delta^2}
  \left( \frac{T}{\pi\hbar D^{\rm sw}} \right)^{d/2}
{\rm Li}_{d/2} \left(-e^{-\frac{\epsilon+\SWgap}{T}}\right),\label{eq:resldt}
\end{equation}
where ${\rm Li}_n(z) = \sum_{k=1}^{\infty} z^k/k^n$ is the polylogarithm. The polylogarithm has the asymptotic behavior
$$
  \mbox{Li}_{d/2}(-e^x) \approx 
  \left\{ \begin{array}{ll} -x^{d/2}/\Gamma(d/2+1) & \mbox{for $x \gg 0$}, \\
  -e^x & \mbox{for $x \ll 0$}, \end{array} \right.
$$
while $\mbox{Li}_{d/2}(-1)$ is a weakly $d$-dependent number of order unity [$\mbox{Li}_{1/2}(-1) \approx -0.605$, $\mbox{Li}_{1}(-1) \approx -0.693$, and $\mbox{Li}_{3/2}(-1) \approx -0.765$].

\begin{figure}[t]
\begin{center}
\includegraphics{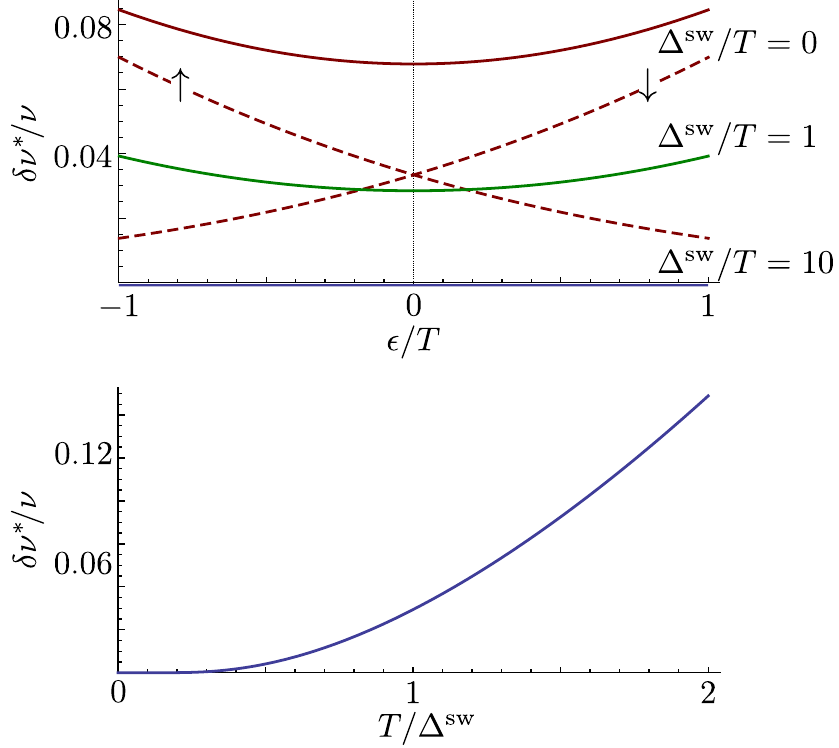}
\caption{(top) The total density-of-states correction $\delta \nu^* = \delta \nu^*_{\ua} + \delta \nu^*_{\da}$ as a function of $\epsilon/T$ for three different ratios $\SWgap/T$ of the spin wave gap $\SWgap$ and the temperature, for a three-dimensional sample. The relative density-of-states correction is given in units of $J^2\bar{s}T^{3/2}/\Delta^2(\hbar D^{\rm sw})^{3/2}$. For $\SWgap=0$ (no spin-wave gap) the corrections for spin-up and spin-down electrons are shown separately (dashed lines). (bottom) The temperature dependence of $\delta\nu^*(0,T)/\nu$ at zero energy, as a function of the ratio $T/\SWgap$ of temperature and spin-wave gap. The relative density-of-states correction is given in units of $J^2\bar{s} (\SWgap)^{3/2}/\Delta^2(\hbar D^{\rm sw})^{3/2}$.\label{fig:dnu3d}}
\end{center}
\end{figure}

The energy and temperature dependence of $\delta \nu^*_{\sigma}$ and of the total density-of-states correction $\delta \nu^* = \delta \nu^*_{\ua} + \delta \nu^*_{\da}$ is shown in Fig.\ \ref{fig:dnu3d} for the case $d=3$. The top panel shows the $\epsilon$-dependence for three representative values of the temperature (larger, equal to, and smaller than the spin-wave gap $\SWgap$). The bottom panel shows the temperature dependence at zero energy.

Expanding Eq.\ (\ref{eq:resldt}) around $\epsilon = 0$ one finds that for small energies, the correction to the total density of states is quadratic in $\epsilon/T$,
\begin{equation}
\frac{\delta \nu^* (\epsilon,T)}{\nu} = \frac{J^2\bar{s}}{2^{d-1}\Delta^2}
  \left( \frac{T}{\pi\hbar D^{\rm sw}} \right)^{d/2}
\left(A_1 + A_2\frac{\epsilon^2}{T^2}\right),
\end{equation}
where we introduced $A_1 = -2 {\rm Li}_{d/2}(-e^{-\SWgap/T})$ and $A_2 = -{\rm Li}_{(d-4)/2}(-e^{-\SWgap/T})$. Both constants reduce to numbers of order unity when the temperature is much larger than the spin-wave gap.

In the limit of (i) very small temperatures or (ii) very far away from the Fermi energy, where $T \ll |\SWgap\pm \epsilon|$, we can use the asymptotic properties of the polylogarithm to find
\begin{align}
\frac{\delta \nu^* (\epsilon,0)}{\nu} = \frac{2^{1-d}J^2\bar{s}}{\Gamma(d/2+1)\Delta^2}
  \left( \frac{|\epsilon|-\SWgap}{\pi\hbar D^{\rm sw}} \right)^{d/2},
\end{align}
if $|\epsilon|>\SWgap$, whereas $\delta \nu(\epsilon,0) = \delta \nu(0,0)$ if $|\epsilon|<\SWgap$.
At low temperatures, the correction to the density of states ceases to be energy dependent for energies which lie closer to the Fermi energy than the spin wave gap energy $\SWgap$. Indeed, at zero temperature the only relevant energy-dependent process is the excitation of spin waves by minority spin electrons or majority spin holes with an energy of at least the spin wave gap away from the Fermi energy.

The temperature dependence of the correction to the density of states also follows from Eq.\ (\ref{eq:resldt}) and the asymptotic dependencies of the polylogarithm. In particular, at the Fermi energy $\epsilon=0$ one finds that $\delta \nu^*$ is proportional to $T^{d/2}$ for temperatures $T$ much larger than the spin-wave gap. The temperature dependence $\delta \nu^*_{\sigma}(0,T) \propto T^{3/2}$ for $d=3$ is the same as the one found in Ref.\ \onlinecite{PhysRevLett.81.705} from a different microscopic mechanism. An important difference with Ref.\ \onlinecite{PhysRevLett.81.705} is that for the mechanism we consider $\delta \nu_{\ua}(0,T) = \delta \nu_{\da}(0,T)$, whereas $\delta \nu_{\ua}(0,T) = - \delta \nu_{\da}(0,T)$ in Ref.\ \onlinecite{PhysRevLett.81.705}.

\subsection{Dirty limit}

We now consider the regime $\Delta\tau_{\rm el}/\hbar \ll 1$ of a dirty ferromagnet. We make the additional assumption that the wave number and energies of all thermal spin waves involved are low enough that $qv_{\rm F}\tau_{\rm el}$, $\omega^{\rm sw}\tau_{\rm el} \ll 1$. This additional assumption allows us to use the diffusion approximation and expand in small $\Delta\tau_{\rm el}/\hbar$, as well as $qv_{\rm F}\tau_{\rm el}$ and $\omega^{\rm sw}\tau_{\rm el}$, which considerably simplifies Eq.\ (\ref{eq:dn}). 

Although our additional assumption $qv_{\rm F}\tau_{\rm el}$, $\omega^{\rm sw}\tau_{\rm el} \ll 1$ is commonly used in the literature,\cite{PhysRevB.79.140408,JPSJ.72.1155,muttalibwoelfle} it poses a rather severe (but not impossible\cite{ferro_dos:footnote2}) restriction on the temperature $T$ and energy $\epsilon$: $\max(T, |\epsilon|) \ll \Delta\hbar^2/(E_{\rm F}\tau_{\rm el})^2$. (The restriction is ``severe'' because $\hbar/E_{\rm F}\tau_{\rm el}$ is the small parameter of the perturbation theory.) The origin of the smallness of allowed temperatures and energies is the large mismatch of the spin wave stiffness and the electronic diffusion constant in the dirty limit, $D/D^{\rm sw} \sim \hbar(k_{\rm F}l_{\rm el})^2/(\Delta\tau_{\rm el}) \gg 1$.

\begin{figure}
\begin{center}
\includegraphics{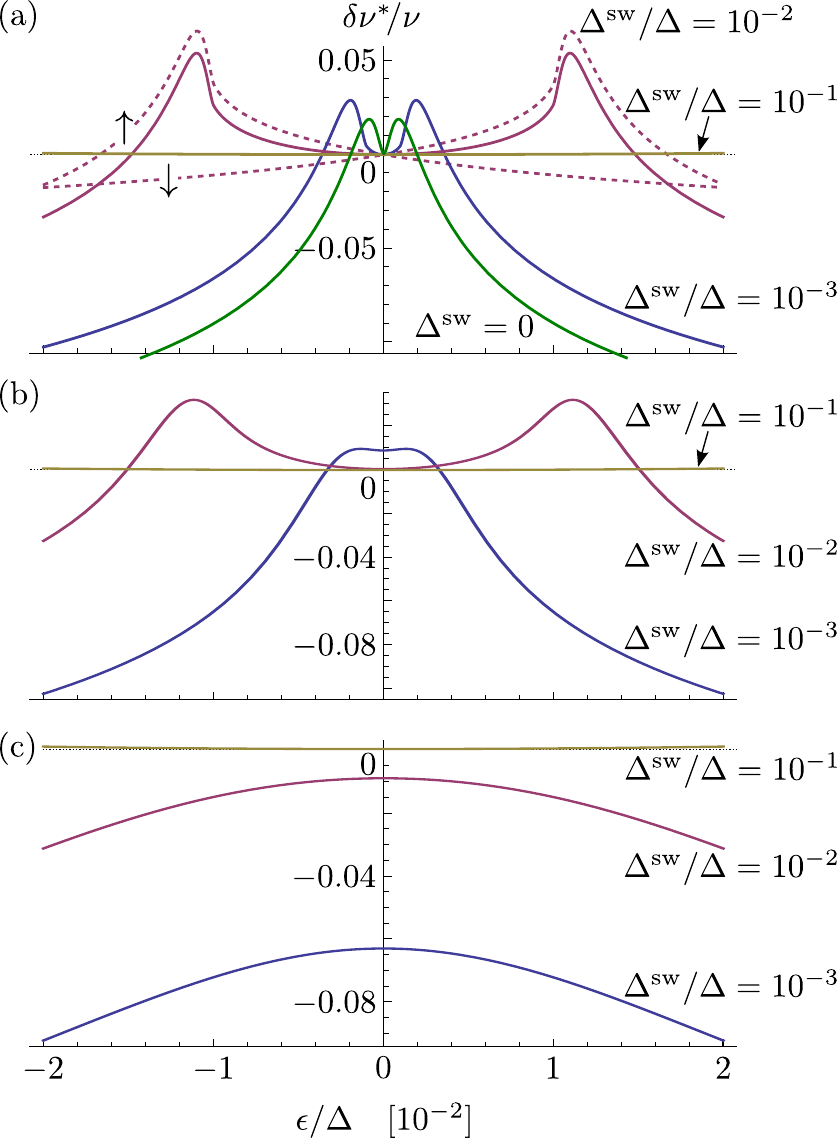}
\caption{Spin-wave induced correction to the density of states for a three-dimensional dirty ferromagnet. We show the correction close to the Fermi energy for three different spin wave gaps $\SWgap$ as indicated in the figure. The temperature was set to (a) $T=0$, (b) $T=10^{-3}\Delta$, and (c) $T=10^{-2}\Delta$. The resulting relative correction $\delta\nu^*(\epsilon,T)/\nu$ is plotted in units of $J^2\bar{s}/(\hbar D)^{3/2}\sqrt{\Delta}$. We have set $D^{\rm sw}/D = 10^{-3}$ in all cases. In (a) we show the corrections at $\SWgap/\Delta = 10^{-2}$ for majority electrons ($\ua$) and minority electrons ($\da$) separately (dashed lines), showing that the peak at negative (positive) energy stems from the correction to the majority (minority) density of states.}\label{fig:plot2}
\end{center}
\end{figure}

\begin{figure}
\begin{center}
\includegraphics{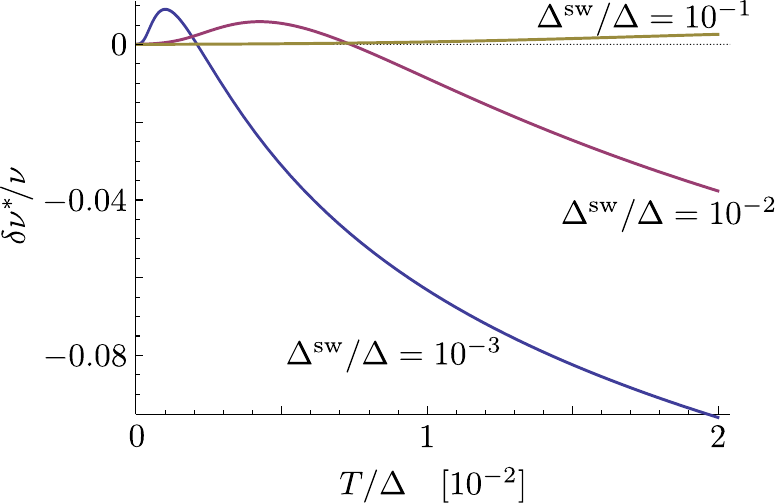}
\caption{Correction to the density of states for a three-dimensional dirty ferromagnet at zero energy as a function of temperature. Curves are shown for three different spin wave gaps $\SWgap$, as indicated in the figure. The relative correction $\delta\nu^*(0,T)/\nu$ is again plotted in units of $J^2\bar{s}/(\hbar D)^{3/2}\sqrt{\Delta}$. We have set $D^{\rm sw}/D = 10^{-3}$ throughout.}\label{fig:plot3}
\end{center}
\end{figure}

In the diffusion approximation, the quantity $1/[1 - \Pi_{\sigma,-\sigma}(\vq,\zeta)]$ appearing in the expression (\ref{eq:dn}) for the density-of-states correction becomes equal to the diffusion propagator,
\begin{eqnarray}
  D(\vq, \zeta) &=&
  \frac{1}{1 - \Pi_{\sigma,-\sigma}(\vq,\zeta)}
  \nonumber \\ &=&
\frac{\hbar}{\tau_{\rm el}(\hbar Dq^2 - i \zeta - i\sigma\Delta)},
\end{eqnarray}
where $D= v_{\rm F}^2\tau_{\rm el}/d$ is the diffusion constant for the conduction electrons in $d$ effective dimensions. For the correction to the density of states this leads to
\begin{align}
\frac{\delta \nu_{\sigma} (\epsilon,T)}{\nu} = -&\frac{J^2\bar{s}}{\pi {\cal V}}\Im \sum_{\vq} \int d\zeta\, \tanh \Big(\frac{\epsilon-\zeta}{2T} \Big) \nonumber\\
& \times \frac{\sigma}{(\zeta + \hbar \sigma\omega_{\vq}^{\rm sw} + i\eta)(\hbar Dq^2 - i \zeta -i \sigma\Delta)^2}.
\label{eq:dosdif}
\end{align}
Although Eq.\ (\ref{eq:dosdif}) can be further evaluated, e.g., in the limit of zero temperature, the resulting expressions are too complex to be insightful, and we therefore prefer a numerical evaluation of the double integral in Eq.\ (\ref{eq:dosdif}) in a few representative limits. Results are shown in Figs.\ \ref{fig:plot2} and \ref{fig:plot3}. In both cases we have set the ratio $D^{\rm sw}/D = 10^{-3}$, consistent with the large mismatch between these two constants mentioned above. 

In Figure \ref{fig:plot2} we show the correction $\delta \nu^*(\epsilon,T)$ for different ratios of the spin-wave gap $\SWgap$ and the exchange splitting $\Delta$ and for different values of the temperature. The figure shows a remarkable non-monotonous dependence on the excitation energy $\epsilon$. This dependence can be most easily understood for the case of zero spin-wave gap and zero temperature ($\SWgap=0$, green line in Fig.\ \ref{fig:plot2}a). At the smallest energies, $|\epsilon| \ll (D^{\rm sw}/D)\Delta$, we find that the maximal wave number contributing to the sum in Eq.\ (\ref{eq:dosdif}) is $q_{\rm max} = \sqrt{|\epsilon|/\hbar D^{\rm sw}} \ll \sqrt{\Delta/\hbar D} $. This implies that for all relevant spin-wave energies we have $\hbar D q^2 \ll \Delta$ leading to $D_{\sigma,-\sigma} (\vq, \zeta) \approx \hbar i /\Delta \tau_{\rm el}$. This means that a typical spin-wave excited electron-hole pair still dephases before it diffuses significantly through the sample, yielding a situation similar to that of the clean ferromagnet treated above. Indeed, if we take $\zeta \ll \Delta$ and $\hbar Dq^2 \ll \Delta$, then Eq.~(\ref{eq:dosdif}) reduces exactly to the clean result (\ref{eq:nusclean}). The correction at small energies then has the same (positive) sign as found before. On the other hand, at large excitation energies $|\epsilon| \gg (D^{\rm sw}/D)\Delta$, we can simplify (\ref{eq:dosdif}) using that for most wave numbers $q$ contributing to the summation we have $D_{\sigma,-\sigma} (\vq, \zeta) \approx 1/ Dq^2 \tau_{\rm el}$, leading to a different sign in (\ref{eq:dosdif}). Indeed, this is a truly diffusive limit, in which the excited electron-hole pairs can diffuse over long distances. The effective spin wave mediated electron-electron interaction $U_{\rm sw}(\vq,i\omega_n) = J^2[\chi_{-+}(\vq,i\omega_n)+\chi_{+-}(\vq,i\omega_n)] < 0$ is of an attractive nature, for which in the diffusive limit the exchange correction is known to yield a correction negative for increasing energy.\cite{aa:book} With a finite spin-wave gap $\SWgap$ the resulting peaks are shifted by $\SWgap$, as seen in Fig.\ \ref{fig:plot2}a). At higher temperatures, as shown in Figs.\ \ref{fig:plot2}b,c, the peaks are smeared out and the correction tends to be a smooth peak centered at the Fermi energy.

In Fig.\ \ref{fig:plot3} we show the temperature dependence of the correction to the DOS at the Fermi level $\epsilon = 0$, using temperatures small enough that the condition $T,|\epsilon| \ll \Delta\hbar^2/(E_{\rm F}\tau_{\rm el})^2$ can be satisfied. The non-monotonous temperature dependence has the same origin as the non-monotonous energy dependence of Fig.\ \ref{fig:plot2}. When the temperature is larger than the spin wave gap, $T\gtrsim \SWgap$, the thermal broadening of the double peak structure around the Fermi level becomes strong enough to reduce $\delta\nu(0,T)$ below the reference value $\delta\nu(0,0)$, which results in a change of sign.

\section{Conclusion}\label{sec:6}

For the spin wave induced correction to the density of states we find qualitatively different results depending on whether $\Delta\tau_{\rm el}/\hbar$ is small or large. For the clean limit $\Delta\tau_{\rm el}/\hbar \gg 1$, which is the most realistic situation, we derived an analytical expression for the density-of-states correction $\delta \nu(\epsilon,T)$ for one-, two-, and three-dimensional samples. At zero excitation energy $\epsilon$ (measured with respect to the Fermi energy) in a $d$-dimensional sample the correction has a power-law temperature dependence $\delta\nu^* \propto T^{d/2}$ for temperatures larger than the spin wave gap $\SWgap$. At $T=0$ and away from the Fermi level, the energy dependence of the correction also follows a power-law, $\delta\nu^* \propto (|\epsilon|-\SWgap)^{d/2}$ for energies larger than $\SWgap$. In the dirty limit $\Delta\tau_{\rm el}/\hbar \ll 1$ we found that the correction to the density of states has a non-monotonous dependence on energy and temperature, with a peak for temperatures or energies near the spin-wave gap $\SWgap$.

Relevant questions to address are how large the correction is in comparison with the correction resulting from electron-electron (Coulomb) interactions in the ferromagnet and how it compares to the mechanism of Ref.\ \onlinecite{PhysRevLett.81.705}. For the first question we restrict ourselves to the experimentally most relevant clean limit $\Delta\tau_{\rm el}/\hbar \gg 1$. In three dimensions the correction due to electron-electron interactions is\cite{aa:book}
\begin{equation}
\delta\nu^* (\epsilon,0)
  = \frac{1}{2\sqrt{2}\pi^2} \frac{\sqrt{|\epsilon|}}{(\hbar D)^{3/2}},
\end{equation}
for very low temperatures and
\begin{equation}
\delta\nu^* (0,T) \approx 0.038\frac{\sqrt{T}}{(\hbar D)^{3/2}},
\end{equation}
at the Fermi level. The corresponding spin-wave-induced corrections found here have have a power-law dependence of $\propto \max(|\epsilon|^{3/2},T^{3/2})$ (neglecting the spin-wave gap $\SWgap$ for simplicity). This means that at small energies and temperatures the correction due to electron-electron interactions dominates, and that there are minimum energy and temperature scales $\epsilon_{\rm min}$ and $T_{\rm min}$ above which the spin wave induced correction can become the dominant one. A straightforward comparison of the two corrections yields
\begin{equation}
\epsilon_{\rm min},\ T_{\rm min} \sim \left( \frac{\sqrt{\Delta\tau_{\rm el}/\hbar}}{E_{\rm F}\tau_{\rm el}/\hbar}\right)^3 E_{\rm F}.
\end{equation}
Since $E_{\rm F} \tau_{\rm el}/\hbar$ is the large parameter of the perturbation theory, this is well within the validity range of the diagrammatic perturbation theory. Using typical parameters for Fe,\cite{ashmer,PhysRevLett.67.2363} we estimate this crossover scale as $\sim 1$~K. We thus expect the temperature dependence of the density-of-states correction to be given by a $T^{3/2}$ power law for temperatures above $T_{\rm min}$.

The density-of-states correction of Ref.\ \onlinecite{PhysRevLett.81.705} is, again to second order in the exchange coupling $J$, 
\begin{equation}
  \delta \nu^*_{\sigma}(\epsilon,T)
  \sim - \frac{J^2 \bar s(\nu_{\sigma} - \nu_{-\sigma})}{\Delta^2 {\cal V}} \sum_{\vq} n_{\rm B}(\hbar \omega_{\vq}^{\rm sw}),
\end{equation} 
where $n_{\rm B}$ is the Bose-Einstein distribution function and $\nu_{\ua}$ and $\nu_{\da}$ are the unperturbed densities of states for majority and minority electrons, respectively. This expression is similar to our result (\ref{eq:nusclean}) for the clean limit, which has the factor $\nu_{\sigma} - \nu_{-\sigma}$ replaced by $\nu$ and the Bose-Einstein factor $n_{\rm B}(\hbar\omega_{\vq}^{\rm sw})$ replaced by the Fermi factor $n_{\rm F}(\epsilon + \hbar\omega_{\vq}^{\rm sw})$. Since typically $n_{\rm B} \gg n_{\rm F}$, we conclude that the magnitude of the correction of Ref.\ \onlinecite{PhysRevLett.81.705} is larger than the correction calculated here in the case of a strong ferromagnet, for which $|\nu_{\ua} - \nu_{\da}| \sim \nu$. Distinguishing the two corrections should still be possible, because of the singular dependence of the correction calculated here on the excitation energy $\epsilon$. (No singular $\epsilon$-dependence is reported in Ref.\ \onlinecite{PhysRevLett.81.705}.)

\acknowledgments

We gratefully acknowledge helpful discussions with Georg Schwiete and Martin Schneider.
This work is supported by the Alexander von Humboldt Foundation in the framework of the Alexander von Humboldt Professorship programme, endowed by the Federal Ministry of Education and Research.


\begin{thebibliography}{10}
\bibitem{ferro_dos:footnote1}
As explained in Ref.\ \onlinecite{PhysRevB.84.224433}, this exchange splitting
  $\Delta$ can differ from $J \bar{s}$ if exchange interactions between
  conduction electrons are taken into account. We will thus not make use of the
  relation $\Delta = 2J \bar{s}$, but keep $\Delta$ as an independent parameter
  instead.
\bibitem{ferro_dos:footnote2}
For a dirty ferromagnet with $E_{\rm F} \sim$~eV and having, e.g., $E_{\rm F}\tau/\hbar \sim 20$ (which then corresponds to an electronic elastic scattering time of the order of $10^{-14}$--$10^{-15}$ s), we find the two constraints mentioned above to be $\Delta \ll 1000$~K and $T,|\epsilon| \ll \Delta/400$. This means that, although on the edge of the regime of validity, the diffusive approximation could work in certain cases for energies and temperatures up to $\sim 1$~K.
\end{thebibliography}
\end{document}